\documentclass[11pt]{iopart}

\usepackage{iopams} 
\usepackage{cite,graphicx,multirow}

\newcommand{\MG}[7]{
G^{3,2}_{3,3} \left(#7\,\Bigg|\, 
\begin{array} {c}
#1,\, #2,\, #3 \\ #4,\, #5,\, #6 \end{array} \right)}

\newcommand{\MGa}[5]{
G^{2,2}_{2,2} \left(#5\Bigg|\, 
\begin{array} {c}
#1,\, #2\\ #3,\,  #4 \end{array} \right)}

\newcommand{\binom}[2]{\Big(\begin{array}{c} #1 \\ #2 \end{array}\Big)}
\newcommand{\dfrac}[2]{\displaystyle\frac{#1}{#2}}

\begin{document}

\title[Exact evaluations of some Meijer G-functions]{Exact evaluations of some Meijer G-functions and probability of all eigenvalues real for product of two Gaussian matrices}

\author{Santosh Kumar}

\address{Department of Physics, Shiv Nadar University, Gautam Buddha Nagar, \\
Uttar Pradesh - 201314, India}
\ead{skumar.physics@gmail.com}
\vspace{10pt}

\begin{abstract}
We provide a proof to a recent conjecture by Forrester (2014, {\it J. Phys. A: Math. Theor.} {\bf 47}, 065202) regarding the algebraic and arithmetic structure of Meijer G-functions which appear in the expression for probability of all eigenvalues real for product of two real Gaussian matrices. In the process we come across several interesting identities involving Meijer G-functions.
\end{abstract}

%
\vspace{2pc}

\section{Introduction}

Product of random matrices and the associated eigenvalue spectra exhibit a number of interesting properties and find concrete applications in varied fields of knowledge.
Their study goes back to as early as 1950's  where the focus was on exploring the behavior of dynamical systems, and the accompanying questions related to stochastic differential equations and Lyapunov exponents~~\cite{Bellman1954,FK1960,Berger1984,BL1985,CPV1993}. In the last few years there has been a revival in interest in their investigation because of their fascinating integrability properties\cite{Osborn2004,Forrester2012,AB2012,AIK2013,AKW2013,AS2013,CLTT2013,ARRS2013,Lakshminarayan2013,Forrester2014,FK2014,KZ2014,KKS2015,AI2015} and identification of new problems where they can be applied, such as random graph states~\cite{CNZ2010}, combinatorics~\cite{PZ2011}, quantum entanglement~\cite{Lakshminarayan2013} and multilayered multiple channel telecommunication~\cite{AIK2013,WZCT2015}.

In a recent study by Lakshminarayan~\cite{Lakshminarayan2013} it was shown that the question concerning optimal quantum entanglement is intimately related to the probability that the product of two real 2-dimensional Gaussian random matrices (real Ginibre matrices) has real eigenvalues. This problem was exactly solved in~\cite{Lakshminarayan2013} and extensive numerical exploration was conducted for the product of arbitrary number of matrices of higher dimensionalities.
A complete analytical solution was provided by Forrester in~\cite{Forrester2014} where he found explicit determinantal expressions for the probability of all eigenvalues real for product of any number of Gaussian matrices of arbitrary dimension. Several related asymptotic results were also derived. In the finite dimensionality case the determinants were found to involve certain Meijer G-functions. Meijer G-functions have also appeared as correlation-kernels in product of Ginibre matrices or of truncated unitary matrices~\cite{AI2015,AB2012,AIK2013,AKW2013,KKS2015}, Bures and Cauchy two-matrix models~\cite{FK2014}, and very recently in the results for product of a Wigner matrix and a Wishart matrix~\cite{Kumar2015}. We give a brief introduction to Meijer G-functions in the next section.

One of the intriguing observations made in~\cite{Forrester2014} involved the Meijer G-function $\MG{5/2-j}{5/2-j}{2}{1}{1+k}{1+k}{1}$ which appeared as the kernel in the case of product of two Gaussian matrices. Based on high-precision numerical computation it was conjectured that these are rational multiples of $\pi^2$ for positive integers $j,k$. We provide here a proof to this conjecture by deriving an exact and simple expression which clearly demonstrates why this is the case. In the process we also come across some interesting identities involving Meijer G-functions.

\section{Exact evaluation of a class of Meijer G-functions}

Meijer G-function is defined in terms of Mellin-Bernes type contour integral~\cite{BE1953,Luke1969}:
\begin{equation}
\fl
G^{m,n}_{p,q} \left(z\,\Bigg|\, 
\begin{array} {c}
a_1,...,a_n,a_{n+1},..., a_p \\ b_1,...,\,b_m,b_{m+1},..., b_q \end{array} \right)=\frac{1}{2\pi i}\int_\mathcal{C}\frac{ds\, z^s\prod_{j=1}^m\Gamma(b_j-s)\prod_{j=1}^n\Gamma(1-a_j+s)}{\prod_{j=m+1}^q\Gamma(1-b_j+s)\prod_{j=n+1}^p\Gamma(a_j-s)},
\end{equation}
where an empty product is interpreted as 1. The integer indices satisfy $0\le m\le q, 0\le n\le p$, and the parameters $a$'s, $b$'s may be real or complex such that no pole of $\prod_{j=1}^m\Gamma(b_j-s)$ coincides with any of the poles of $\prod_{j=1}^n\Gamma(1-a_j+s)$. Also, the contour $\mathcal{C}$ is such that it separates the poles of $\prod_{j=1}^m\Gamma(b_j-s)$ from those of $\prod_{j=1}^n\Gamma(1-a_j+s)$; see ~\cite{BE1953,Luke1969} for details. One of the fascinating properties of the family of Meijer G-functions is its closure property under differentiation as well as indefinite integration. Moreover, majority of the established elementary functions and special functions can be represented in terms of Meijer G-functions. A few examples are
\begin{equation}
e^z=G^{1,0}_{0,1} \left(-z\,\Bigg|\, 
\begin{array} {c}
- \\ 0 \end{array} \right),
\end{equation}
\begin{equation}
\ln (1+z)=G^{1,2}_{2,2} \left(z\,\Bigg|\, 
\begin{array} {c}
1,~1\\ 1,~0 \end{array}\right),
\end{equation}
\begin{equation}
\cos z=\sqrt{\pi}\,G^{1,0}_{0,2} \left(\frac{z^2}{4}\,\Bigg|\, 
\begin{array} {c}
-\\ 0,~1/2 \end{array}\right),
\end{equation}
\begin{equation}
\label{IncGamma}
\Gamma(\alpha, z)=G^{2,0}_{1,2} \left(z\,\Bigg|\, 
\begin{array} {c}
1\\ \alpha,~0 \end{array}\right),
\end{equation}
\begin{equation}
\label{Bessel}
J_\nu(z)=G^{1,0}_{0,2} \left(\frac{z^2}{4}\,\Bigg|\, 
\begin{array} {c}
-\\ \nu/2,-\nu/2 \end{array}\right),~~-\frac{\pi}{2}<\mathrm{arg} (z)\leq \frac{\pi}{2},
\end{equation}
\begin{equation}
\label{Elliptic}
K(z)=\frac{1}{2}G^{1,2}_{2,2} \left(-z\,\Bigg|\, 
\begin{array} {c}
1/2,~1/2\\ 0,~0 \end{array}\right).
\end{equation}
The special functions on left hand side of Eqs.~\eref{IncGamma}-\eref{Elliptic} are upper incomplete Gamma function, Bessel function of the first kind, and complete elliptic integral of the first kind, respectively.

For our purpose we begin with the identity~\cite{BE1953,Luke1969}
\begin{equation}
\label{MG1}
\MG{a_1}{a_2}{c}{b_1}{b_2}{c}{z}=\MGa{a_1}{a_2}{b_1}{b_2}{z}.
\end{equation}
The above result follows because of the presence of common parameter $c$ in the upper and lower sets of parameters in the Meijer G-function on the left-hand side~\cite{BE1953,Luke1969,Wolfram}.
The Meijer G-function in the above equation has a representation in terms of regularized Gauss hypergeometric function~\cite{Wolfram}:
\begin{eqnarray}
\fl
\label{MG2}
\nonumber
\MGa{a_1}{a_2}{b_1}{b_2}{z}=\Gamma(1-a_1+b_1)\Gamma(1-a_2+b_1)\Gamma(1-a_1+b_2)\Gamma(1-a_2+b_2)\\
\times z^{b_1}\,_2 \widetilde{F}_1(1-a_1+b_1,1-a_2+b_1;2-a_1-a_2+b_1+b_2;1-z).
\end{eqnarray}
We note that the regularized Gauss hypergeometric function is related to the usual Gauss hypergeometric function by the simple relation $\,_2\widetilde{F}_1(\alpha_1,\alpha_2;\beta;z)=\,_2F_1(\alpha_1,\alpha_2;\beta;z)/\Gamma(\beta)$.
We now use one of the derivative identities for Meijer G-function~\cite{Wolfram},
\begin{equation}
\frac{d}{dz}\left[z^{-c}\MG{a_1}{a_2}{c}{b_1}{b_2}{c}{z}\right]=-z^{-1-c}\,\MG{a_1}{a_2}{c}{b_1}{b_2}{c+1}{z},
\end{equation}
which leads, when applied recursively, to the following expression:
\begin{eqnarray}
\nonumber
\MG{a_1}{a_2}{c}{b_1}{b_2}{c+n}{z}&=(-1)^n z^{c+n}\frac{d^n}{dz^n} \left[z^{-c}\MG{a_1}{a_2}{c}{b_1}{b_2}{c}{z}\right]\\
&=(-1)^n z^{c+n}\frac{d^n}{dz^n} \left[z^{-c}\MGa{a_1}{a_2}{b_1}{b_2}{z}\right].
\end{eqnarray}
This is an interesting result which can be directly used along with~\eref{MG2} to demonstrate the truth of the conjecture in~\cite{Forrester2014}. 
However, we make further progress and insert relation~\eref{MG2} in the above equation and use the following results to carry out the $n$-fold differentiation:
\begin{equation}
\label{D1}
\frac{d^n}{dz^n}(f(z) g(z))=\sum_{\mu=0}^{n} \binom{n}{\mu} \frac{d^{n-\mu}f(z)}{dz^{n-\mu}}\frac{d^\mu g(z)}{dz^\mu},
\end{equation}
\begin{equation}
\label{D2}
\frac{d^m}{dz^m}z^{-\alpha}=(-1)^m (\alpha)_m z^{-\alpha-m},
\end{equation}
\begin{equation}
\label{D3}
\frac{d^m}{dz^m}\,_2 \widetilde{F}_1(\alpha,\beta;\gamma;1-z)=(-1)^m(\alpha)_m (\beta)_m\,_2 \widetilde{F}_1(\alpha+m,\beta+m;\gamma+m;1-z).
\end{equation}
Here $\binom{n}{\mu}$ represents the binomial coefficient and $(a)_n=\Gamma(a+n)/\Gamma(a)$ is the Pochhammer symbol.
Incorporating all these results and carrying out some simplification, we arrive at the following identity:
\begin{eqnarray}
\label{key1}
\fl
\nonumber
\MG{a_1}{a_2}{c}{b_1}{b_2}{c+n}{z}=\Gamma(1-a_1+b_2)\Gamma(1-a_2+b_2) \\
\fl
\nonumber
\times\sum_{\mu=0}^{n} \binom{n}{\mu} (c-b_1)_{n-\mu}\,\Gamma(\mu+1-a_1+b_1)\Gamma(\mu+1-a_2+b_1)\\
\fl
\times z^{\mu+b_1}\,_2 \widetilde{F}_1(\mu+1-a_1+b_1,\mu+1-a_2+b_1;\mu+2-a_1-a_2+b_1+b_2;1-z).
\end{eqnarray}
Substituting  $z=1$ and using
\begin{equation}
\,_2 \widetilde{F}_1(\alpha,\beta;\gamma;0)=\frac{1}{\Gamma(\gamma)},
\end{equation}
we obtain
\begin{eqnarray}
\label{key2}
\nonumber
\MG{a_1}{a_2}{c}{b_1}{b_2}{c+n}{1}=\Gamma(1-a_1+b_2)\Gamma(1-a_2+b_2)\\
\times\sum_{\mu=0}^{n} \binom{n}{\mu}(c-b_1)_{n-\mu}\, \frac{\Gamma(\mu+1-a_1+b_1)\Gamma(\mu+1-a_2+b_1)}{\Gamma(\mu+2-a_1-a_2+b_1+b_2)}.
\end{eqnarray}
Equations~\eref{key1} and~\eref{key2} are two key contributions of this work. We use~\eref{key2} in the next section to prove the result predicted in~\cite{Forrester2014}.

\section{Probability of all eigenvalues real for product of two real Gaussian matrices}

Let us consider two $N\times N$-dimensional real Gaussian matrices $X$ and $Y$ which have the associated joint probability measure
\begin{equation}
\label{JPD}
\mathcal{P}(X,Y)d[X] d[Y]=\left(\frac{1}{2\pi}\right)^{N^2}e^{-\frac{1}{2}\tr(XX^T+YY^T)}d[X] d[Y].$$
\end{equation}
Here tr represents trace, and T stands for transpose. Also $d[X]$ refers to the product of differentials of all matrix elements of $X$. Similar definition is to be understood for $d[Y]$.
Forrester has derived an exact determinantal expression for the probability of all $N$ eigenvalues real for product of $m$ number of $N\times N$-dimensional standard real Gaussian matrices~\cite{Forrester2014}. We quote here the result for $m=2$, i.e., the probability of all eigenvalues being real for the product $XY$, where $X$ and $Y$ are from~\eref{JPD}. For even $N$ we have
\begin{eqnarray}
\label{Pev}
\fl
p_{N,N}^{XY}=\left(\prod_{j=1}^N\frac{1}{\Gamma^2(j/2)}\right)\det\left[\MG{5/2-j}{5/2-j}{2}{1}{1+k}{1+k}{1}\right]_{j,k=1,...,N/2},
\end{eqnarray}
while for odd $N$ we have
\begin{eqnarray}
\label{Pod}
\fl
\nonumber
p_{N,N}^{XY}=\left(\prod_{j=1}^N\frac{1}{\Gamma^2(j/2)}\right)\\
\fl
\times\det\left[\left[\MG{5/2-j}{5/2-j}{2}{1}{1+k}{1+k}{1}\right]_{j=1,...,(N+1)/2\atop k=1,...,(N-1)/2} \left[\Gamma^2(j-1/2)\right]_{j=1,...,(N+1)/2}\right].
\end{eqnarray}
Using high-precision numerical computation it was conjectured in~\cite{Forrester2014} that the Meijer G-functions in the above expressions lead to values which are rational multiples of $\pi^2$. We show below that this is indeed the case.

We set $a_1=a_2=5/2-j, b_1=1, b_2=k+1,c=2, n=k-1 $ in \eref{key2}, which gives on little simplification the following remarkable identity
\begin{eqnarray}
\label{key}
\nonumber
\MG{5/2-j}{5/2-j}{2}{1}{1+k}{1+k}{1}\\
\nonumber
=\Gamma(k)\Gamma^2(j+k-1/2)\sum_{\mu=0}^{k-1}\frac{\Gamma^2(\mu+j-1/2)}{\Gamma(\mu+1)\Gamma(\mu+2j+k-1)}\\
=\pi^2\frac{\Gamma(k)\Gamma^2(2j+2k-1)}{\Gamma^2(j+k)}\sum_{\mu=0}^{k-1}\frac{16^{2-\mu-2j-k}\Gamma^2(2\mu+2j-1)}{\Gamma(\mu+1)\Gamma^2(\mu+j)\Gamma(\mu+2j+k-1)}.
\end{eqnarray}
This is the central result of this paper which confirms the algebraic and arithmetic structure predicted in~\cite{Forrester2014}. Using Eq.~\eref{key}, it takes less than a second to evaluate the above Meijer G-function with $j,k$ values as large as 1000 in Mathematica~\cite{Mathematica}. We have, for $j,k=1,...,5$,
\renewcommand\arraystretch{1.4}
\begin{eqnarray}
\fl
\nonumber
\left[\MG{5/2-j}{5/2-j}{2}{1}{1+k}{1+k}{1}\right]_{j,k=1,...,5}\\
\fl
=\pi ^2\left(
\begin{array}{ccccc}
 \frac{1}{2^2} & \frac{39 }{2^7} & \frac{10335 }{2^{13}} & \frac{2997855 }{2^{18}} & \frac{6149253915 }{2^{25}} \\ 
 \frac{3 }{2^7} & \frac{435 }{2^{13}} & \frac{72555 }{2^{18}} & \frac{91686735 }{2^{25}} & \frac{48462643845 }{2^{30}} \\
 \frac{135 }{2^{13}} & \frac{16695 }{2^{18}} & \frac{15107715 }{2^{25}} & \frac{5645015145 }{2^{30}} & \frac{6504362819955 }{2^{36}} \\
 \frac{7875 }{2^{18}} & \frac{6024375 }{2^{25}} & \frac{1840070925 }{2^{30}} & \frac{1683904397175 }{2^{36}} & \frac{1105018317277875 }{2^{41}} \\
 \frac{3472875 }{2^{25}} & \frac{955040625 }{2^{30}} & \frac{768670177275 }{2^{36}} & \frac{432899597505375 }{2^{41}} & \frac{2645687420488987875 }{2^{49}} \\
\end{array}
\right).
\end{eqnarray}
\begin{table}
\renewcommand{\arraystretch}{2}
\caption{Exact values and numerical values (6 significant digits for $N>1$) for probability $p_{N,N}^{XY}$ of all eigenvalues real for product of two $N\times N$-dimensional real Gaussian matrices are listed in the second and third columns, respectively. The fourth column displays numerical values of the the ratio $(4/\pi)(p_{N-1,N-1}^{XY}p_{N+1,N+1}^{XY})/(p_{N,N}^{XY})^2$ and supports the leading large $N$ form $(\pi/4)^{N^2/2}$. }
\centering
\begin{tabular}{|c|c|c|c| }
\hline
 \multirow{2}{*}{$N$} & \multicolumn{2}{|c|}{$p_{N,N}^{XY}$}  & \multirow{2}{*}{{\footnotesize $\dfrac{4}{\pi}\dfrac{p_{N-1,N-1}^{XY}p_{N+1,N+1}^{XY}}{(p_{N,N}^{XY})^2}$}}  \\ \cline{2-3}
  &Exact  & Numerical value &  \\
\hline\hline
1 & 1 & 1 & $-$ \\
\hline
2 & $\dfrac{\pi}{2^2}$ & $7.85398\times 10^{-1}$ & 1.01321 \\
\hline
3 & $\dfrac{5\pi}{2^5}$ & $4.90874 \times 10^{-1}$ &  1.00500 \\
\hline
4 & $\dfrac{201\pi^2}{2^{13}}$ & $2.42162 \times 10^{-1}$ & 1.00446\\
\hline
5 & $\dfrac{10013\pi^2}{2^{20}}$ & $9.42462 \times 10^{-2}$ & 1.00257\\
\hline
6 & $\dfrac{64011585\pi^3}{2^{36}}$ & $2.88821 \times 10^{-2}$ & 1.00229\\
\hline
7 & $\dfrac{31625532537\pi^3}{2^{47}}$ & $6.96751 \times 10^{-3} $ & 1.00156 \\
\hline
8 & $\dfrac{8012440011007425\pi^4}{2^{69}}$ & $1.32219 \times 10^{-3}$ & 1.00142\\
\hline
9 & $\dfrac{39186641315011126281\pi^4}{2^{84}}$ & $1.97341 \times 10^{-4}$ & 1.00106\\
\hline
10 & $\dfrac{6286653393344610981261954345\pi^5}{2^{116}}$ & $2.31574  \times 10^{-5}$ & 1.00098\\
\hline
11 & $\dfrac{304070790487188921741594082108725\pi^5}{2^{135}}$ & $2.13636 \times 10^{-6} $ & 1.00077\\
\hline
\end{tabular}
\label{TabWigCorr}
\end{table}
Interestingly, Mathematica also returns the sum in~\eref{key} in terms of a generalized hypergeometric function, which gives
\begin{eqnarray}
\fl
\nonumber
\MG{5/2-j}{5/2-j}{2}{1}{1+k}{1+k}{1}=\Gamma^2(j-1/2)\Gamma^2(k)-\Gamma(k)\Gamma^4(j+k-1/2)\\
\fl
\times\,_3\widetilde{F}_2(1,j+k-1/2,j+k-1/2;k+1,2j+2k-1;1).
\end{eqnarray}
This result, when combined with~\eref{key}, leads to yet another interesting identity
\begin{eqnarray}
\fl
\nonumber
\,_3\widetilde{F}_2(1,j+k-1/2,j+k-1/2;k+1,2j+2k-1;1)=\frac{\Gamma^2(j-1/2)\Gamma(k)}{\Gamma^4(j+k-1/2)}\\
-\frac{1}{\Gamma^2(j+k-1/2)}\sum_{\mu=0}^{k-1}\frac{\Gamma^2(\mu+j-1/2)}{\Gamma(\mu+1)\Gamma(\mu+2j+k-1)}.
\end{eqnarray} 
We note here that $\,_3\widetilde{F}_2(\alpha_1,\alpha_2,\alpha_3;\beta_1,\beta_2;z)=\,_3F_2(\alpha_1,\alpha_2,\alpha_3;\beta_1,\beta_2;z)/(\Gamma(\beta_1)\Gamma(\beta_2))$.

\begin{figure}[t!]
\centering
 \includegraphics[width=0.8\textwidth]{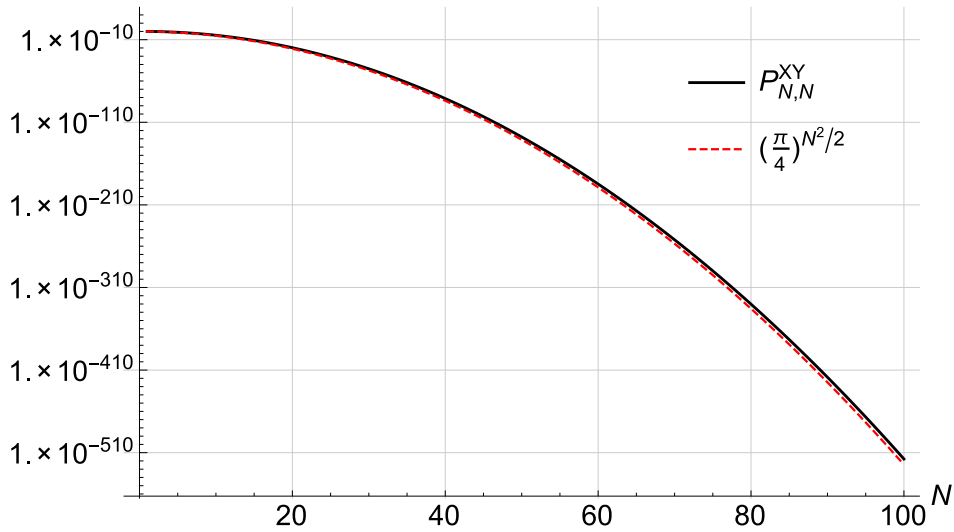}
\caption{Plots of $P_{N,N}^{XY}$ and the leading large-$N$ asymptotic $(\pi/4)^{N^2/2}$ for $N$ up to 100. The $y$-axis scale is logarithmic.}
\label{PXY}
\end{figure}

With the result~\eref{key} at our disposal, we can also calculate exact results for $p_{N,N}^{XY}$ using~\eref{Pev} and~\eref{Pod} as rational multiples of powers of $\pi$. Exact expressions for this probability for $N$ as large as 100 is obtained in about 5 seconds using Mathematica. We list exact results up to $N=11 $ in Table 1, along with the numerical values with six significant digits. Also displayed in the third column are the numerical values of the ratio $(4/\pi)(p_{N-1,N-1}^{XY}p_{N+1,N+1}^{XY})/(p_{N,N}^{XY})^2$, which corroborates the large $N$ leading result $(\pi/4)^{N^2/2}$~\cite{Forrester2014}. In Fig.~\ref{PXY} we show the plots of exact $P_{N,N}^{XY}$ and the large $N$ leading contribution $(\pi/4)^{N^2/2}$. The curves are in good agreement.

\section{Summary and Outlook}
In this work we explored some identities involving Meijer G and hypergeometric functions. Using these we proved a recent conjecture of Forrester regarding the algebraic and arithmetic structure of Meijer G-function appearing in the study of probability of all eigenvalues real for product of two real Gaussian random matrices.
Whether similar structures exist in the case of product of more than two matrices, remains an open question.

\ack The author is grateful to Prof. P. J. Forrester for fruitful correspondence. 


\end{document}